\documentclass[aps,pre,showkeys,
longbibliography,
reprint,                
twocolumn,              
nofootinbib,
floatfix]{revtex4-1}

\usepackage{amsmath}
\usepackage{amsfonts}
\usepackage{graphicx}
\usepackage{tikz}
\usepackage{subcaption}

\usepackage{psfrag}

\usepackage[colorlinks=true,hyperfootnotes=true,breaklinks=true]{hyperref}
\usepackage[capitalise]{cleveref}

\begin{document}

\title{Percolation thresholds on triangular lattice for neighbourhoods containing sites up-to the fifth coordination zone}

\author{\href{http://home.agh.edu.pl/malarz/}{Krzysztof Malarz}}
\thanks{\includegraphics[width=10pt]{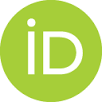}~\href{https://orcid.org/0000-0001-9980-0363}{0000-0001-9980-0363}}
\email{malarz@agh.edu.pl}
\affiliation{\href{http://www.agh.edu.pl/en}{AGH University of Science and Technology},
\href{http://www.pacs.agh.edu.pl/indexe.html}{Faculty of Physics and Applied Computer Science},\\
al. Mickiewicza 30, 30-059 Krak\'ow, Poland}

\begin{abstract}
We determine thresholds $p_c$ for random-site percolation on a triangular lattice for all available neighborhoods containing sites from the first to the fifth coordination zones, including their complex combinations. 
There are 31 distinct neighbourhoods. The dependence of the value of the percolation thresholds $p_c$ on 
the coordination number $z$ are tested against various theoretical predictions.
The newly proposed single scalar index $\xi=\sum_i z_ir_i^2/i$ (depending on the 
coordination zone number $i$, the neighbourhood coordination number $z$ and the 
square-distance $r^2$ to sites in $i$-th coordination zone from the central site) allows to
differentiate among various neighbourhoods and relate $p_c$ to $\xi$. The thresholds roughly follow a power 
law $p_c\propto\xi^{-\gamma}$ with $\gamma\approx 0.710(19)$.  
\end{abstract}

\date{\today}

\maketitle

\section{Introduction}

The concepts of site and bond percolation \cite{bookDS,Wierman2014} introduced in middle fifties \cite{Broadbent1957,Hammersley1957} and since then have been applied in various fields of science ranging from agriculture \cite{ISI:000518460000003} via studies of polymer composites \cite{ISI:000528948100007}, materials science \cite{ISI:000514848600043}, forest fires \cite{Kaczanowska2002}, oil and gas exploration \cite{ISI:000524118200031}, quantifying urban areas \cite{ISI:000523958600016}, Bitcoins transfer \cite{Bartolucci2020}, diseases propagation \cite{2101.00550} to transportation networks \cite{ISI:000528691800009} (see Refs.~\onlinecite{Li_2021,Saberi2015} for reviews).

Percolation is an example of phenomenon where a geometrical phase transition (on $d$-dimensional lattice) takes place. The critical parameter (called percolation threshold $p_c$ \cite{Frisch1961}) separates two phases: one
for a low occupation probability $p<p_c$ and the other for $p>p_c$. In the low-$p$ phase the system behaves as 
an insulator (without connectivity path leading between system boundaries) while for the high-$p$ phase 
there is a giant component spanning the system and connecting opposite boundaries; in effect the system behaves as a
conductor, when one refers to the electric analogy.

The percolation thresholds were initially estimated for nearest neighbour interactions  \cite{Dean_1963,Dean_Bird_1967} but later also complex (or extended) neighbourhoods were studied for 2D (square \cite{Dalton_1964,Domb1966,Gouker1983,Galam2005a,Galam2005b,Majewski2007,2010.02895}, triangular \cite{Dalton_1964,Domb1966,Iribarne1999,2006.15621}, honeycomb \cite{Dalton_1964}), 3D (simple cubic \cite{Kurzawski2012,Malarz2015,2010.02895}) and 4D (simple hyper-cubic \cite{1803.09504}) lattices.

\begin{figure*}
\begin{subfigure}[b]{0.16\textwidth}
\caption{\label{fig:1nn}}
\begin{tikzpicture}[scale=0.39]
\clip (-3.5,-3.5) rectangle (3.5,3.5);
\begin{scope}
\pgftransformcm{1}{0}{1/2}{sqrt(3)/2}{\pgfpoint{0cm}{0cm}} 
\draw[style=help lines] (-4,-4) grid[step=1] (4,4);
\end{scope}
\begin{scope}
\pgftransformcm{1}{0}{-1/2}{sqrt(3)/2}{\pgfpoint{0cm}{0cm}} 
\draw[style=help lines] (-4,-4) grid[step=1] (4,4);
\end{scope}
\filldraw[red] (0,0) circle (7pt);
\draw[ultra thick,orange] (0,0) circle (1);
\filldraw (  1,0) circle (7pt);
\filldraw ( -1,0) circle (7pt);
\filldraw ( .5,1.732050807/2) circle (7pt);
\filldraw (-.5,1.732050807/2) circle (7pt);
\filldraw ( .5,-1.732050807/2) circle (7pt);
\filldraw (-.5,-1.732050807/2) circle (7pt);
\end{tikzpicture}
\end{subfigure}
\hfill 
\begin{subfigure}[b]{0.16\textwidth}
\caption{\label{fig:2nn}}
\begin{tikzpicture}[scale=0.39]
\clip (-3.5,-3.5) rectangle (3.5,3.5);
\begin{scope}
\pgftransformcm{1}{0}{1/2}{sqrt(3)/2}{\pgfpoint{0cm}{0cm}}
\draw[style=help lines] (-4,-4) grid[step=1] (4,4);
\end{scope}
\begin{scope}
\pgftransformcm{1}{0}{-1/2}{sqrt(3)/2}{\pgfpoint{0cm}{0cm}} 
\draw[style=help lines] (-4,-4) grid[step=1] (4,4);
\end{scope}
\filldraw[red] (0,0) circle (7pt);
\draw[ultra thick,orange] (0,0) circle (1.732050807);
\filldraw (  0, 1.732050807) circle (7pt);
\filldraw (  0,-1.732050807) circle (7pt);
\filldraw ( 1.5,1.732050807/2) circle (7pt);
\filldraw (-1.5,1.732050807/2) circle (7pt);
\filldraw ( 1.5,-1.732050807/2) circle (7pt);
\filldraw (-1.5,-1.732050807/2) circle (7pt);
\end{tikzpicture}
\end{subfigure}
\hfill 
\begin{subfigure}[b]{0.16\textwidth}
\caption{\label{fig:3nn}}
\begin{tikzpicture}[scale=0.39]
\clip (-3.5,-3.5) rectangle (3.5,3.5);
\begin{scope}
\pgftransformcm{1}{0}{1/2}{sqrt(3)/2}{\pgfpoint{0cm}{0cm}}
\draw[style=help lines] (-4,-4) grid[step=1] (4,4);
\end{scope}
\begin{scope}
\pgftransformcm{1}{0}{-1/2}{sqrt(3)/2}{\pgfpoint{0cm}{0cm}} 
\draw[style=help lines] (-4,-4) grid[step=1] (4,4);
\end{scope}
\filldraw[red] (0,0) circle (7pt);
\draw[ultra thick,orange] (0,0) circle (2);
\filldraw ( 2,0) circle (7pt);
\filldraw (-2,0) circle (7pt);
\filldraw ( 1,1.732050807) circle (7pt);
\filldraw (-1,1.732050807) circle (7pt);
\filldraw ( 1,-1.732050807) circle (7pt);
\filldraw (-1,-1.732050807) circle (7pt);
\end{tikzpicture}
\end{subfigure}
\hfill 
\begin{subfigure}[b]{0.16\textwidth}
\caption{\label{fig:4nn}}
\begin{tikzpicture}[scale=0.39]
\clip (-3.5,-3.5) rectangle (3.5,3.5); 
\begin{scope} 
\pgftransformcm{1}{0}{1/2}{sqrt(3)/2}{\pgfpoint{0cm}{0cm}} 
\draw[style=help lines] (-4,-4) grid[step=1] (4,4);
\end{scope}
\begin{scope}
\pgftransformcm{1}{0}{-1/2}{sqrt(3)/2}{\pgfpoint{0cm}{0cm}} 
\draw[style=help lines] (-4,-4) grid[step=1] (4,4);
\end{scope}
\filldraw[red] (0,0) circle (7pt);
\draw[ultra thick,orange] (0,0) circle (3*1.732050807/2);
\filldraw ( 2,1.732050807) circle (7pt);
\filldraw (-2,1.732050807) circle (7pt);
\filldraw ( 2,-1.732050807) circle (7pt);
\filldraw (-2,-1.732050807) circle (7pt);
\filldraw ( 2.5,-1.732050807/2) circle (7pt);
\filldraw (-2.5,-1.732050807/2) circle (7pt);
\filldraw ( 2.5,1.732050807/2) circle (7pt);
\filldraw (-2.5,1.732050807/2) circle (7pt);
\filldraw ( 0.5,-3*1.732050807/2) circle (7pt);
\filldraw (-0.5,-3*1.732050807/2) circle (7pt);
\filldraw ( 0.5,3*1.732050807/2) circle (7pt);
\filldraw (-0.5,3*1.732050807/2) circle (7pt);
\end{tikzpicture}
\end{subfigure}
\hfill 
\begin{subfigure}[b]{0.16\textwidth}
\caption{\label{fig:5nn}}
\begin{tikzpicture}[scale=0.39]\clip (-3.5,-3.5) rectangle (3.5,3.5); 
\begin{scope} 
\pgftransformcm{1}{0}{1/2}{sqrt(3)/2}{\pgfpoint{0cm}{0cm}} 
\draw[style=help lines] (-4,-4) grid[step=1] (4,4);
\end{scope}
\begin{scope}
\pgftransformcm{1}{0}{-1/2}{sqrt(3)/2}{\pgfpoint{0cm}{0cm}} 
\draw[style=help lines] (-4,-4) grid[step=1] (4,4);
\end{scope}
\filldraw[red] (0,0) circle (7pt);
\draw[ultra thick,orange] (0,0) circle (3);
\filldraw ( 3,0) circle (7pt);
\filldraw (-3,0) circle (7pt);
\filldraw ( 1.5, 3*1.732050807/2) circle (7pt);
\filldraw ( 1.5,-3*1.732050807/2) circle (7pt);
\filldraw (-1.5, 3*1.732050807/2) circle (7pt);
\filldraw (-1.5,-3*1.732050807/2) circle (7pt);
\end{tikzpicture}
\end{subfigure}
\caption{\label{fig:neighbors}(Color online). Basic neighbourhoods corresponding to subsequent coordination zones $i=1,\cdots,5$ on the triangular lattice. The symbol $r$ stands for the Euclidean distance of black sites to the central one
(in red) and $z$ indicates the number of sites in the neighbourhood. Examples of fifteen neighborhoods containing the next-next-next-next-nearest neighbors (with sites from the 5-th coordination zone) on triangular lattice are presented in Figure A1 in Ref.~\onlinecite{SupMat}.
(a) \textsc{tr}-1: $i=1$, $r^2=1$, $z=6$,
(b) \textsc{tr}-2: $i=2$, $r^2=3$, $z=6$,
(c) \textsc{tr}-3: $i=3$, $r^2=4$, $z=6$,
(d) \textsc{tr}-4: $i=4$, $r^2=7$, $z=12$,
(e) \textsc{tr}-5: $i=5$, $r^2=9$, $z=6$.}
\end{figure*}
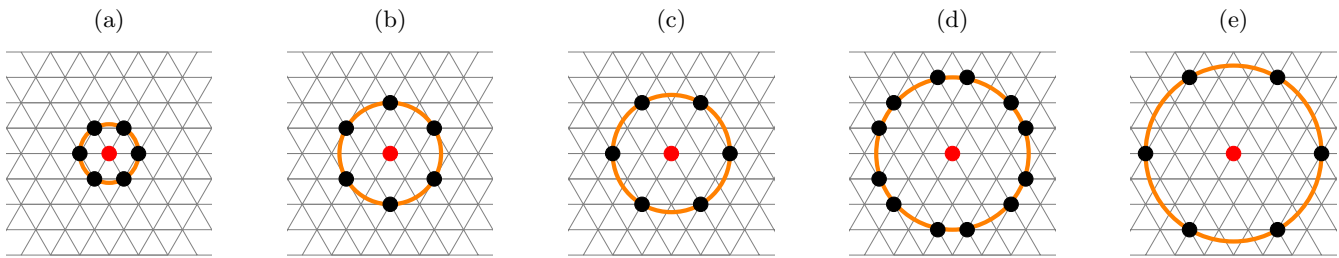

Very recently, we have computed percolation thresholds for random site percolation on triangular lattice with complex neighbourhoods with hexagonal symmetry \cite{2006.15621}. Here we supplement these results with 31 percolation thresholds estimations for all neighbourhoods on triangular lattice containing sites from the first, the second, the third, the fourth and the fifth coordination zones (see \Cref{fig:neighbors}). 
Some of these neighbourhoods---those containing sites from the fifth coordination zone---are presented in Figure A1 in Ref.~\onlinecite{SupMat}.
The lattice names follow convention proposed in Ref.~\onlinecite{2010.02895} reflecting lattice topology (here \textsc{tr}, i.e. triangular lattice) and numerical string specifying the coordination zones, where sites constituting the neighbourhood come from.

Additionally---for triangular lattice and complex neighbourhoods---we test the dependence of $p_c$ on
the coordination number $z$, following the idea of Ref.~\onlinecite{2010.02895}. However, instead of values of the percolation thresholds for selected (mainly compact) neighbourhoods, we use the mean values $\bar p_c$ 
of percolation thresholds $p_c$ averaged over all available neighbourhoods with given coordination number $z$. 
Unfortunately, values of $\bar p_c$ do not follow any of dependencies proposed in Ref.~\onlinecite{2010.02895}.

Finally, we propose a scalar quantity $\xi$ which may be helpful for differentiating among various neighbourhoods. The quantity is based on the coordination zone $i$, the sites number $z$ and the sites distances $r$ to the central site in the neighbourhood. The dependency of $p_c$ on this newly proposed index $\xi$ follows roughly a power law with an exponent close to $-0.710(19)$.
 
\section{\label{sec:methods}Methods}

\begin{figure*}
\begin{subfigure}[b]{0.45\textwidth}
\psfrag{Smax/N}{$\langle\mathcal{S}_{\max}\rangle/N$}
\psfrag{n/N}{$n/N$}
\psfrag{L}{$L=$}
\caption{\label{fig:2a}}
\includegraphics[width=0.95\textwidth]{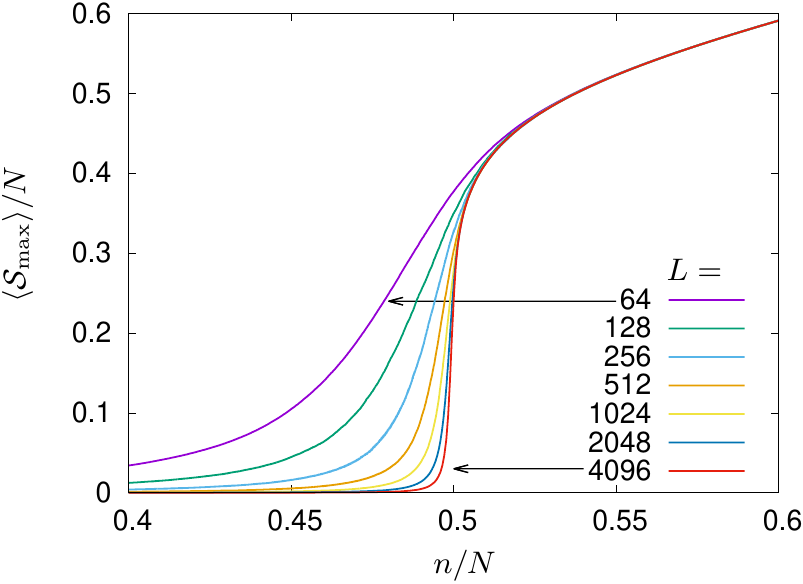}
\end{subfigure}
\begin{subfigure}[b]{0.45\textwidth}
\psfrag{PmaxLx}{$\mathcal{P}_{\max}\cdot L^{\beta/\nu}$}
\psfrag{p}{$p$}
\psfrag{L}{$L=$}
\caption{\label{fig:2b}}
\includegraphics[width=0.95\textwidth]{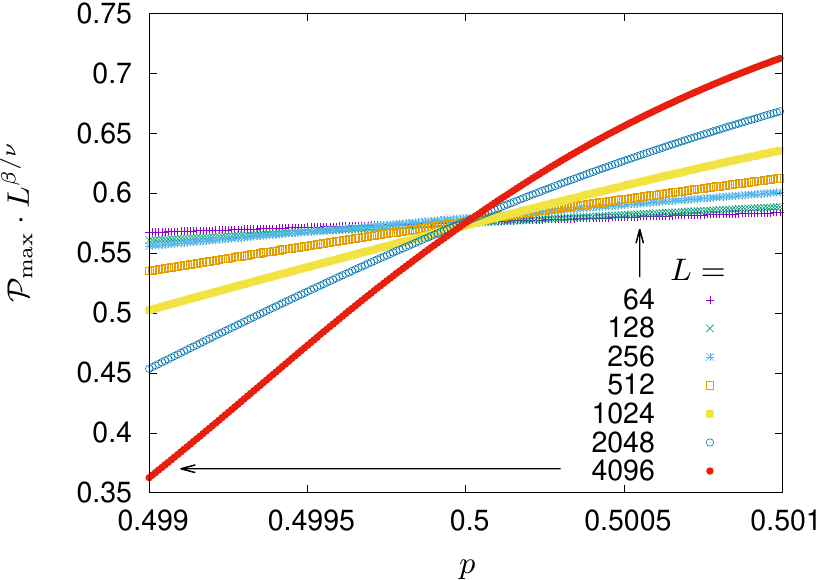}
\end{subfigure}
\begin{subfigure}[b]{0.45\textwidth}
\psfrag{lambda}{$\lambda$}
\psfrag{p}{$p$}
\psfrag{L}{$L=$}
\caption{\label{fig:2c}}
\includegraphics[width=0.95\textwidth]{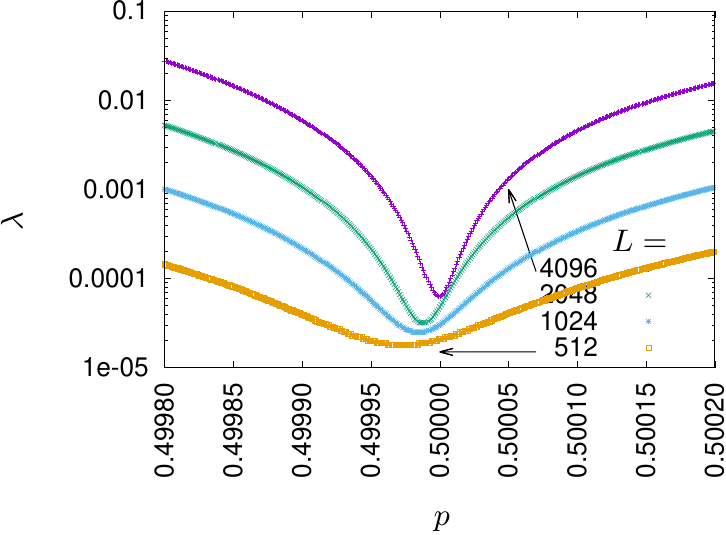}
\end{subfigure}
\begin{subfigure}[b]{0.45\textwidth}
\psfrag{L^{-1/nu}}{$L^{-1/\nu}$}
\psfrag{hat pc}{$\hat p_c$}
\caption{\label{fig:2d}}
\includegraphics[width=0.95\textwidth]{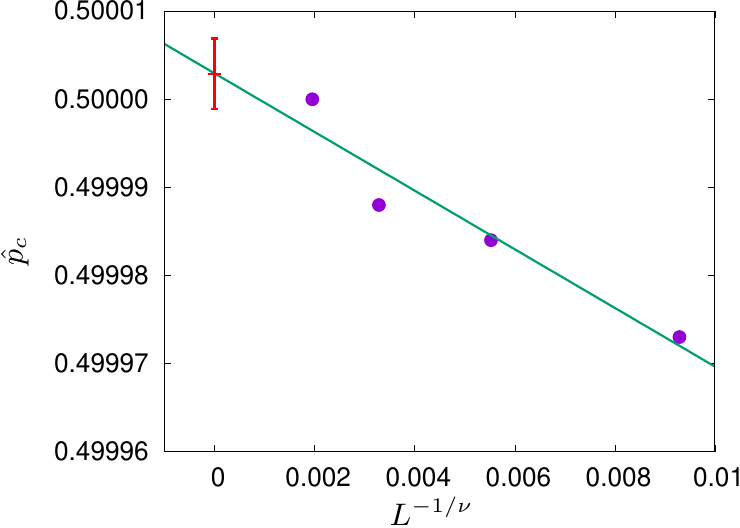}
\end{subfigure}
\caption{\label{fig:2}(Color online). Subsequent steps of the percolation threshold estimation (example for \textsc{tr}-5 lattice).
(a) The mean largest cluster size $\langle\mathcal{S}_{\max}\rangle$ vs. the number of occupied sites, normalized to the system size $N$.
(b) The probability that an arbitrarily chosen site belongs to the largest
cluster
$\mathcal{P}_{\max}(p;L)\cdot L^{\beta/\nu}$ on the occupation probability $p$.
(c) The dependence of $\lambda(p)$ on the occupation probability $p$. The minima give estimates of the percolation thresholds $\hat p_c$.
(d) Finite size-scaling corrections to $\hat p_c$ vs. $L^{-1/\nu}$ according to \Cref{eq:pcvsL}.}
\end{figure*}

In order to evaluate the percolation thresholds $p_c$ we follow the scheme applied previously in Ref.~\onlinecite{2006.15621}.
Namely, we combine \citet{NewmanZiff2001}, \citet{Bastas2014} algorithms with finite size corrections \cite{bookDS,bookVP} to estimate $p_c$:
\begin{enumerate}
\item Based on the \citet{NewmanZiff2001} algorithm we calculate the size of the largest cluster $\langle\mathcal{S}_{\max}(n;N)\rangle$ on the number of occupied sites $n$ (\Cref{fig:2a}). The brackets $\langle\cdots\rangle$ represent averaging over $R=10^5$ lattice realizations. Applying a Gaussian
approximation to the Bernoulli distribution 
\begin{equation}
\label{eq:binom}
\mathcal{B}(n;N,p)=\binom{N}{n}p^n(1-p)^{N-n}
\end{equation}
one can calculate 
\begin{equation}
\mathcal{S}_{\max}(p;N)=\sum_{n=1}^N\langle\mathcal{S}_{\max}(n;N)\rangle\mathcal{B}(n;N,p)
\end{equation}
for different values of the occupation probability $p$ using the Gauss function:
\begin{equation}
\label{eq:gauss}
\mathcal{B}(n;N,p)\approx
\mathcal{G}(n;\mu,\sigma) = \frac{1}{\sqrt{2\pi\sigma^2}} \exp\left( -\frac{(n-\mu)^2}{2\sigma^2} \right),
\end{equation}
with the expected value $\mu=pN$ and variance $\sigma^2=p(1-p)N$. 
The dependence $\mathcal{S}_{\max}(p;N)$ yields a probability that an arbitrarily chosen site belongs to the largest
cluster
\begin{equation}
\label{eq:Pmax}
\mathcal{P}_{\max}(p;L)=\mathcal{S}_{\max}(p;L)/N,
\end{equation}
for system with $N=L^2$ sites and the linear system size $L=64$, 128, 256, 512, 1024, 2048 and 4096 as presented in \Cref{fig:2b};
\item Using \citet{Bastas2014} algorithm we minimize the pair-wise difference 
\begin{equation}
\label{eq:lambda}
\lambda(p) = \sum_{i\ne j} [\mathcal{H}(p;N_i)-\mathcal{H}(p;N_j)]^2
\end{equation}
function (see \Cref{fig:2c}) with $\mathcal{H}(p;L) = L^{\beta/\nu}\cdot\mathcal{P}_{\max}(p;L)+1/[L^{\beta/\nu}\cdot\mathcal{P}_{\max}(p;L)]$ \cite{PhysRevE.84.066112} and with exponents $\beta=\frac{5}{36}$ and $\nu=\frac{4}{3}$ \cite[p.~54]{bookDS}. The minimum of $\lambda(p)$ estimates the percolation threshold $\hat p_c$;
\item Finally, in \Cref{fig:2d} we plot the estimated values of percolation thresholds $\hat p_c(L)$ for different ranges of summation in \Cref{eq:lambda}---up to $L=\max(L_{i,j})=512$, $1024$, $2048$ and $4096$.
According to the standard finite size scaling \cite[p.~77]{bookDS} 
\begin{equation}
\label{eq:pcvsL}
\hat p_c(L) = p_c + a\cdot L^{-1/\nu},
\end{equation}
where $p_c$ is the percolation threshold for an infinitely large system. The least squares linear fit to data presented in \Cref{fig:2d} predicts $p_c$ and its uncertainty $u(p_c)$. 
\end{enumerate}

\begin{table*}
\caption{\label{tab:pc}Estimated values of random site triangular lattice percolation thresholds $p_c$ for various complex neighbourhoods. The lattice name encodes the coordination zones $i$, to which sites in the neighbourhood belong. Also the coordination number $z$, the index $\xi$ and the value of $p$ at the minimum of $\lambda$ are presented.}
\begin{ruledtabular}
\begin{tabular}{lrrlllr}
              lattice 	& $z$ & $\xi$ 	& $p$ at $\min(\lambda)$   & $p_c$                & $p_c$                  & earlier estimations \\
                	    &	   &	& $(128\le L\le 4096)$ 	                & $(128\le L\le 4096)$ & $(64\le L\le 4096)$    &  \\
                	    &	   &	& with $\Delta p=10^{-6}$               &                      &	                    &  \\ \hline
                	    
\textsc{tr}-1,2,3,4,5 	&	36	& 54.8& 0.1157~40\footnote{$L\le 2048$} & 0.1157370(74)\footnotemark[1] & 0.1157399(58)\footnotemark[1] & 0.115847(21) \cite{2006.15621}\\
\textsc{tr}-2,3,4,5   	&	30	& 48.8& 0.1174~40\footnotemark[1] & 0.117467(15)\footnotemark[1]  & 0.117460(10)\footnotemark[1] & 0.117579(41) \cite{2006.15621}\\
\textsc{tr}-1,3,4,5   	&	30	& 45.8& 0.1215~48\footnotemark[1] & 0.121583(14)\footnotemark[1]  & 0.1215730(83)\footnotemark[1] \\	
\textsc{tr}-1,2,4,5   	&	30	& 46.8& 0.1225~93\footnotemark[1] & 0.1226215(68)\footnotemark[1] & 0.1226119(30)\footnotemark[1] \\	
\textsc{tr}-1,2,3,5   	&	24	& 33.8& 0.1522~59	& 0.152297(17)  & 0.152282(10)  \\
\textsc{tr}-3,4,5     	&	24	& 39.8& 0.1255~48	& 0.1255511(43) & 0.1255483(43) \\
\textsc{tr}-2,4,5     	&	24	& 40.8& 0.1266~22	& 0.126653(11)  & 0.1266400(40) \\
\textsc{tr}-2,3,5     	&	18	& 27.8& 0.1616~45	& 0.161664(15)  & 0.161653(15)  \\
\textsc{tr}-1,4,5     	&	24	& 37.8& 0.1316~69 & 0.1316677(13) & 0.13166484(66)& 0.131792(58) \cite{2006.15621} \\
\textsc{tr}-1,3,5     	&	18	& 24.8& 0.1700~42	& 0.1700473(87) & 0.170039(10)  \\
\textsc{tr}-1,2,5     	&	18	& 25.8& 0.1762~42	& 0.176263(11)  & 0.1762610(92) \\
\textsc{tr}-4,5       	&	18	& 31.8& 0.1402~43 & 0.1402453(79) & 0.1402382(92) & 0.140286(5) \cite{2006.15621} \\
\textsc{tr}-3,5       	&	12	& 18.8& 0.1957~03	& 0.1956981(14) & 0.1957039(24) \\
\textsc{tr}-2,5\footnote{equivalent of \textsc{TR}-1,2}       	&	12	& 19.8& 0.2902~68	& 0.290280(20)  & 0.290279(17)  \\
\textsc{tr}-1,5       	&	12	& 16.8& 0.2095~62	& 0.209563(13)  & 0.209561(10)  \\
\textsc{tr}-5\footnote{equivalent of \textsc{TR}-1}         	&	6	& 10.8& 0.5000~00 & 0.5000029(40) & 0.4999961(55) & $\frac{1}{2}$ \cite{2006.15621} \\[1mm]\hline\hline

\textsc{tr}-1,2,3,4   	&	30	&  44& 0.1358~13\footnotemark[1] & 0.135817(29)\footnotemark[1] & 0.135823(27)\footnotemark[1] \\ 
\textsc{tr}-2,3,4     	&	24	&  38& 0.1391~15	& 0.1391118(33) & 0.1391117(26) \\
\textsc{tr}-1,3,4     	&	24	&  35& 0.1443~07	& 0.1443064(38) & 0.1443074(32) \\
\textsc{tr}-1,2,4     	&	24	&  36& 0.1489~78	& 0.1489791(88) & 0.1489757(74) \\
\textsc{tr}-3,4       	&	18	&  29& 0.1519~32	& 0.1519532(26) & 0.1519393(35) \\
\textsc{tr}-2,4       	&	18	&  30& 0.1584~53	& 0.1584634(54) & 0.1584620(43) \\
\textsc{tr}-1,4       	&	18	&  27& 0.1651~88	& 0.165186(14)  & 0.165186(12)  \\
\textsc{tr}-4         	&	12	&  21& 0.1924~37 & 0.1924356(68) & 0.1924428(50) & 0.192410(43) \cite{2006.15621} \\[1mm]\hline\hline

\textsc{tr}-1,2,3     	&	18	&  23& 0.2154~62 & 0.21546261(91)& 0.2154657(17) & 0.215484(19) \cite{2006.15621}, 0.215 \cite{Iribarne1999} \\
\textsc{tr}-2,3       	&	12	&  17& 0.2320~12 & 0.232019(23)  & 0.232020(20)  & 0.232008(38) \cite{2006.15621} \\
\textsc{tr}-1,3       	&	12	&  14& 0.2645~25	& 0.264545(25)  & 0.264539(21)  \\
\textsc{tr}-3\footnotemark[3]         	&	6	&  8& 0.5000~24 & 0.500027(31)  & 0.500013(23)  & $\frac{1}{2}$ \cite{2006.15621} \\[1mm]\hline\hline

\textsc{tr}-1,2       	&	12	&  15& 0.2902~67 & 0.290261(22)  & 0.290258(19)  & 0.295 \cite{Dalton_1964} \\
\textsc{tr}-2\footnotemark[3]         	&	6	&  9& 0.4999~85 & 0.499987(20)  & 0.499978(20)  & $\frac{1}{2}$ \cite{2006.15621} \\[1mm]\hline\hline

\textsc{tr}-1         	&	6	&   6& 0.4999~93 & 0.499994(17)  & 0.499996(14)  & $\frac{1}{2}$ \cite[p. 17]{bookDS} \\

\end{tabular}
\end{ruledtabular}
\end{table*}

\section{\label{sec:results}Results}

In \Cref{fig:2} we present examples of results used to predict the percolation thresholds $p_c$ as described in \Cref{sec:methods}.  

In \Cref{fig:2a} the dependence of $\langle\mathcal{S}_{\max}(n;L)\rangle$ on the number of occupied sites $n$ (normalized to the system size $N$) are presented. The dependence for all discussed neighbourhoods is presented in Figure A2 in Ref.~\onlinecite{SupMat}. With the increase of the system size $N=L^2$ the curves describing this dependence become steeper and steeper.
For infinitely large systems the function $d\langle\mathcal{S}_{\max}(n;L)\rangle/dn$ becomes discontinuous at $p=p_c$. 

In \Cref{fig:2b} we show the dependence of the normalized probability of a randomly chosen site belonging to the largest cluster $\mathcal{P}_{\max}(p;L)\cdot L^{\beta/\nu}$ on the sites occupation probability $p$ for various linear system sizes $L$.
The curves representing this dependence for all discussed neighbourhoods are presented in Figure A3 in Ref.~\onlinecite{SupMat}.
The abscissas of the points where curves intercept each other estimate the percolation thresholds $\hat p_c$.

In \Cref{fig:2c} the dependence of $\lambda(p)$ on the occupation probability $p$ is presented.
The curves representing this dependence for all discussed neighbourhoods are presented in Figure A4 in Ref.~\onlinecite{SupMat}.
The minima of $\lambda(p)$ give estimates of the percolation thresholds $\hat p_c$.

In \Cref{fig:2d} finite size corrections to $\hat p_c$ according to \Cref{eq:pcvsL} are presented.
The curves representing this dependence for all discussed neighbourhoods are presented in Figure A5 in Ref.~\onlinecite{SupMat}.
The initial value of the linear fit function predicts the percolation threshold values $p_c$.  

The obtained percolation thresholds $p_c$ together with their uncertainties and earlier estimates are gathered in \Cref{tab:pc}.
We used all lattice sizes $64\le L\le 4096$ for the estimation of $p_c$ (they are collected in the sixth column of \Cref{tab:pc}), while estimations given in the fifth column rely on systems with $128\le L\le 4096$ (i.e. the smallest systems, for $L=64$, have been excluded from calculations).
For neighbourhoods containing 30 or more sites we were able to carry out simulations for lattices up to $L=2048$.
To check how the minimisation of pair-wise difference function $\lambda$ \eqref{eq:lambda} reduces the finite size effects we present also values of $p$ for which $\lambda$ reaches the minimum when $p$ is scanned with $\Delta p=10^{-6}$ accuracy (see the fourth column of \Cref{tab:pc}).

\section{\label{S:discussion}Discussion}

\begin{figure*}
\begin{subfigure}[b]{0.49\textwidth}
\psfrag{pc}{$p_c$}
\psfrag{z}{$z$}
\caption{\label{fig:3a}}
\includegraphics[width=.99\textwidth]{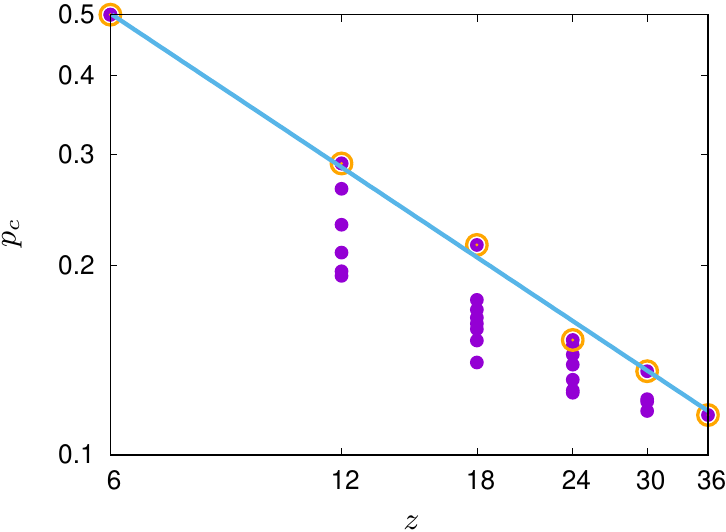}
\end{subfigure}
\begin{subfigure}[b]{0.49\textwidth}
\psfrag{xi}{$\xi$}
\psfrag{pc}{$p_c$}
\psfrag{TR-25}{\textsc{tr}-2,5}
\caption{\label{fig:3b}}
\includegraphics[width=.99\textwidth]{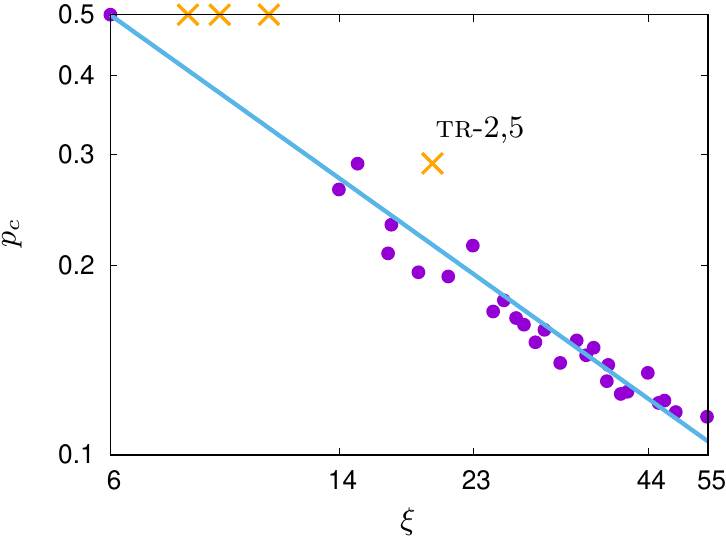}
\end{subfigure}
\caption{\label{fig:3}(Color online). Percolation thresholds for complex neighbourhoods on triangular lattice. 
(a) Degenerated dependence of the percolation threshold $p_c$ on the coordination number $z$. The maxima of $p_c$ for fixed $z$ follow \Cref{eq:maxpcvsz}. The power is $\delta\approx 0.811$. (b) Power fit \eqref{eq:fitpcvsxi} of the percolation thresholds $p_c$ vs. $\xi$ for complex neighbourhoods. The exponent $\gamma$ in \Cref{eq:maxpcvsz} is $0.710(19)$. The $p_c$ values for equivalents of \textsc{tr}-1 and \textsc{tr}-1,2 neighbourhoods (marked with crosses) are excluded from fitting.}
\end{figure*}

In Ref.~\onlinecite{2010.02895} several analytical 
formulas for the dependence of the percolation threshold $p_c$ on the coordination number $z$ were tested, with 
\begin{equation} \label{eq:Ziff}
p_c=c/(z+b), 
\end{equation}
\begin{equation} \label{eq:Koza} 
p_c=1-\exp(d/z)
\end{equation}
among the others.
These formulas work fine for ``compact'' neighbourhoods (for instance \textsc{tr}-1, \textsc{tr}-1,2,3 and \textsc{tr}-1,2,3,4,5, here).

Unfortunately, for complex neighbourhoods these formulas must fail as the dependence $p_c$ on $z$ is ``degenerated'', i.e. several values of $p_c$ are associated with the same number $z$ of sites in the neighbourhood (see \Cref{fig:3a}, and also Figure 4b in Ref.~\onlinecite{Majewski2007} for the square lattice).
These degeneration is also observed for basic neighbourhoods what brought some brickbats \cite{ISI:A1997WD54600080} on the possibility of existing universal formulas for the percolation threshold (depending solely on the spatial dimension $d$ of the system and the coordination number $z$), as proposed by \citet{PhysRevE.53.2177}.   

To remove this degeneracy we tested formulas \eqref{eq:Ziff} and \eqref{eq:Koza} for the mean values $\bar p_c$ of percolation thresholds. The averaging, denoted by the bar, goes over percolation thresholds for neighbourhoods for fixed number $z$. For instance $\bar p_c(z=6)$ is the mean value of $p_c$ for \textsc{tr}-1, \textsc{tr}-2, \textsc{tr}-3, \textsc{tr}-5 lattices, while $\bar p_c(z=30)$ is based on values of $p_c$ for \textsc{tr}-1,2,3,4, \textsc{tr}-1,2,4,5, \textsc{tr}-1,3,4,5 and \textsc{tr}-2,3,4,5 lattices, respectively.
Additionally, we checked the quality of such fits for the Galam--Mauger formula \cite{PhysRevE.53.2177}, which for fixed network topology reduces to the following power-law dependence 
\begin{equation} \label{eq:Galam-Mauger} 
p_c\propto (z-1)^{-a}.
\end{equation}
Unfortunately, the formulas $\bar p_c(z)$ describing the dependence on $z$ are not consistent with either \eqref{eq:Ziff} or \eqref{eq:Koza} or \eqref{eq:Galam-Mauger}.
However, we notice that the maximum value of $p_c$ for the fixed value of coordination number $z$ follows a power law
\begin{equation} \label{eq:maxpcvsz}
    \max_{z=\text{const}} p_c\propto z^{-\delta}
\end{equation}
with the power $\delta\approx 0.811(20)$ (see \Cref{fig:3a}).

As the combination of $d$ and $z$ seems to be insufficient to differentiate among various lattices with assumed neighbourhoods we are looking for another index $\xi$ which may be useful for both: deriving a formula $p_c(\xi)$ and removing the $p_c$-degeneracy.
The attempts for such index identification were undertaken also in graph theory (for topological invariants for trees \cite{Piec2005}) and in organic chemistry (for molecular topological index \cite{Gutman_1994,Schultz_1989,Wiener_1947}).

The primary reason why the formula \eqref{eq:Galam-Mauger} does not work in our case seems to be related to the fact that it does not take into account spatial distribution of neighbors.
We expect some corrections depending on the distance of neighbors from the central node. 
To implement this basic intuition, we found, using trial and error method, that a relatively simple heuristic formula
\begin{equation} \label{eq:xi} 
\xi=\sum_i z_i r_i^2/ i
\end{equation}
does the job, in the sense that the percolation thresholds for complex neighborhoods in the studied model are indeed very well described by the scaling
\begin{equation} \label{eq:fitpcvsxi} 
p_c\propto\xi^{-\gamma}
\end{equation}
with $\gamma\approx 0.710(19)$, except of a single value of $p_c$ for \textsc{tr}-2,5 (see \Cref{fig:3b}). 
It remains to check in other models of this type how universal is the scaling of the percolation thresholds in the effective weighted coordination number $\xi$ \eqref{eq:xi}.

The values of $\xi$ are presented in \Cref{tab:pc}.

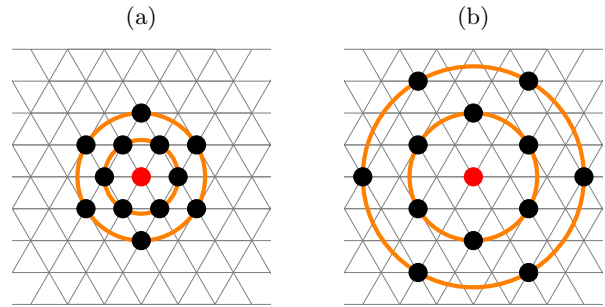
\begin{figure}
\begin{subfigure}[b]{0.23\textwidth}
\caption{\label{fig:tr-1}}
\begin{tikzpicture}[scale=0.49]
\clip (-3.5,-3.5) rectangle (3.5,3.5);
\begin{scope}
\pgftransformcm{1}{0}{1/2}{sqrt(3)/2}{\pgfpoint{0cm}{0cm}} 
\draw[style=help lines] (-4,-4) grid[step=1] (4,4);
\end{scope}
\begin{scope}
\pgftransformcm{1}{0}{-1/2}{sqrt(3)/2}{\pgfpoint{0cm}{0cm}} 
\draw[style=help lines] (-4,-4) grid[step=1] (4,4);
\end{scope}
\filldraw[red] (0,0) circle (7pt);
\draw[ultra thick,orange] (0,0) circle (1);
\draw[ultra thick,orange] (0,0) circle (1.732050807);
\filldraw (  1,0) circle (7pt);
\filldraw ( -1,0) circle (7pt);
\filldraw ( .5,1.732050807/2) circle (7pt);
\filldraw (-.5,1.732050807/2) circle (7pt);
\filldraw ( .5,-1.732050807/2) circle (7pt);
\filldraw (-.5,-1.732050807/2) circle (7pt);
\filldraw (   0, 1.732050807) circle (7pt);
\filldraw (   0,-1.732050807) circle (7pt);
\filldraw ( 1.5,1.732050807/2) circle (7pt);
\filldraw (-1.5,1.732050807/2) circle (7pt);
\filldraw ( 1.5,-1.732050807/2) circle (7pt);
\filldraw (-1.5,-1.732050807/2) circle (7pt);
\end{tikzpicture}
\end{subfigure}
\hfill 
\begin{subfigure}[b]{0.23\textwidth}
\caption{\label{fig:tr-25}}
\begin{tikzpicture}[scale=0.49]
\clip (-3.5,-3.5) rectangle (3.5,3.5);
\begin{scope}
\pgftransformcm{1}{0}{1/2}{sqrt(3)/2}{\pgfpoint{0cm}{0cm}} 
\draw[style=help lines] (-4,-4) grid[step=1] (4,4);
\end{scope}
\begin{scope}
\pgftransformcm{1}{0}{-1/2}{sqrt(3)/2}{\pgfpoint{0cm}{0cm}} 
\draw[style=help lines] (-4,-4) grid[step=1] (4,4);
\end{scope}
\filldraw[red] (0,0) circle (7pt);
\draw[ultra thick,orange] (0,0) circle (3);
\draw[ultra thick,orange] (0,0) circle (1.732050807);
\filldraw (   0, 1.732050807) circle (7pt);
\filldraw (   0,-1.732050807) circle (7pt);
\filldraw ( 1.5,1.732050807/2) circle (7pt);
\filldraw (-1.5,1.732050807/2) circle (7pt);
\filldraw ( 1.5,-1.732050807/2) circle (7pt);
\filldraw (-1.5,-1.732050807/2) circle (7pt);
\filldraw (   3,0) circle (7pt);
\filldraw (  -3,0) circle (7pt);
\filldraw ( 1.5, 3*1.732050807/2) circle (7pt);
\filldraw ( 1.5,-3*1.732050807/2) circle (7pt);
\filldraw (-1.5, 3*1.732050807/2) circle (7pt);
\filldraw (-1.5,-3*1.732050807/2) circle (7pt);

\end{tikzpicture}
\end{subfigure}
\caption{\label{fig:equiv12=25}(Color online). Topological equivalence of the neighbourhoods \textsc{tr}-1,2 and \textsc{tr}-2,5.
The distances to the central site in \textsc{tr}-2,5 are $\sqrt{3}$ times longer than in \textsc{tr}-1,2 lattice.
(a) \textsc{tr}-1,2: $r^2=1$ and 3. 
(b) \textsc{tr}-2,5: $r^2=3$ and 9.}
\end{figure}

Please note that $p_c(\textsc{tr}\text{-1})=p_c(\textsc{tr}\text{-2})=p_c(\textsc{tr}\text{-3})=p_c(\textsc{tr}\text{-5})$ are exactly the same and equal to $\frac{1}{2}$ as these neighbourhoods are equivalent to each other as described in Ref.~\onlinecite{2006.15621}.   
Thus $p_c$ values for \textsc{tr}-2, \textsc{tr}-3, \textsc{tr}-5 neighbourhoods are excluded from fitting. 
The same occurs for \textsc{tr}-2,5 which is equivalent to \textsc{tr}-1,2 as presented in \Cref{fig:equiv12=25}.
In general, in case of neighbourhood equivalence the system should
be characterized by $\xi$ value associated with this neighbourhood
which has the smallest indices that characterize the connection.

\section{\label{S:conclusions}Conclusions}

Concluding, in this paper we estimated percolation thresholds $p_c$ for random site triangular lattice percolation and for neighborhoods containing sites from the first to fifth coordination zone.
The estimated values of percolation thresholds are collected in \Cref{tab:pc}.

We note that the method \citet{Bastas2014} allows (at least partially) to get rid of finite size effects. The minima of the pairwise difference $\lambda$ function presented in \Cref{tab:pc} are consistent with a five digit accuracy with the percolation threshold $p_c$ obtained for infinite lattice according to \Cref{eq:pcvsL}. The five digit accuracy seems to outperform by at least one order of magnitude practical requirements on experimenters in any field of science where percolation theory may be applied.

Percolation thresholds for extended neighbourhoods were utilised in studies ranging from 
agriculture \cite{ISI:000518460000003}, 
materials chemistry \cite{Alguero_2020},
magnetic \cite{PhysRevB.97.165121,ISI:000400959000004,PhysRevB.94.054407}
and electronic \cite{ISI:000419615800018} properties of solids,
nanoengineering \cite{Xu_2014}, etc.
These applications utilised results for square or simple cubic lattices. However, also finding percolation thresholds for extended neighbourhoods on honeycomb lattice may be attractive for better understanding some phenomena as proposed for instance in Ref.~\onlinecite{PhysRevB.73.054422}.
Thus application of the results for triangular lattice presented here cannot be excluded.

As the percolation thresholds $p_c(z)$ are multiply degenerated, we propose the weighted square distance $\xi$ to differentiate among various neighbourhoods. This index seems to be effective in this respect (at least for neighbourhoods investigated here).

Finally, the $p_c(\xi)$ dependence follows roughly a power law \eqref{eq:fitpcvsxi}. 
The obtained results may be useful in further searching for the universal formula on percolation thresholds \cite{Lebrecht_2021}.


\bibliography{percolation,km,this}


\end{document}